\documentclass[12pt]{iopart}

\usepackage{graphicx}
\usepackage{iopams}

\newtheorem{hypothesis}{\sc Hypothesis}

\begin{document}


\title{Dimension dependent energy thresholds for discrete breathers}

\author{Michael Kastner\dag }

\address{\dag\ Physikalisches Institut, Lehrstuhl f\"ur Theoretische Physik I, Universit\"at Bayreuth, 95440 
Bayreuth, Germany}

\ead{Michael.Kastner@uni-bayreuth.de}

\begin{abstract}
Discrete breathers are time-periodic, spatially localized solutions of the equations of motion for a system of 
classical degrees of freedom interacting on a lattice. We study the existence of energy thresholds for discrete 
breathers, i.e., the question whether, in a certain system, discrete breathers of arbitrarily low energy exist, 
or a threshold has to be overcome in order to excite a discrete breather. Breather energies are found to have a 
positive lower bound if the lattice dimension $d$ is greater than or equal to a certain critical value $d_c$, 
whereas no energy threshold is observed for $d<d_c$. The critical dimension $d_c$ is system dependent and can be 
computed explicitly, taking on values between zero and infinity. Three classes of Hamiltonian systems are 
distinguished, being characterized by different mechanisms effecting the existence (or non-existence) of an 
energy threshold.
\end{abstract}


\section{Introduction}

Discrete breathers are time-periodic, spatially localized solutions of the equations of motion for a system of 
classical degrees of freedom interacting on a lattice. They are also called {\em intrinsically localized}, in 
distinction to Anderson localization triggered by disorder. A necessary condition for their existence is the 
nonlinearity of the equations of motion of the system, and the existence of discrete breathers has been proved 
rigorously for some classes of systems \cite{MacAub,Bambusi,LiSpiMac,AuKoKa,James1,James2,JaNo}. In contrast to 
their analogues in continuous systems, the existence of discrete breathers is a generic phenomenon, which 
accounts for considerable interest in these objects from a physical point of view in the last decade. In fact, 
recent experiments could demonstrate the existence of discrete breathers in various real systems such as 
low-dimensional crystals \cite{Swanson_ea}, antiferromagnetic materials \cite{SchwarzEnSie}, Josephson junction 
arrays \cite{TriMaOr,Binder_ea}, molecular crystals \cite{EdHamm}, coupled optical waveguides \cite{Mandelik_ea}, 
and micromechanical cantilever arrays \cite{Sato_ea}.

Properties of discrete breathers, as well as of some generalizations of discrete breathers, have been studied in 
detail in a large variety of different models. However, apart from existence proofs and studies of the spatial 
localization of discrete breathers (which is typically exponential), only a few general results exist. Among 
these it is worth mentioning the remarkable result by Flach, Kladko, and MacKay \cite{FlaKlaMac} on energy 
thresholds for discrete breathers in one-, two-, and three-dimensional lattices (and subsequent generalizations 
to systems with long range interactions \cite{Flach2} and to partially isochronous potentials \cite{DoFla}). 
Results on energy thresholds have practical relevance, as they can assist in choosing a proper energy range for 
the detection of discrete breathers in real experiments or computer experiments. It is found that, for 
Hamiltonian systems of infinite size, energy thresholds depend on the spatial dimension $d$ of the system. A 
critical spatial dimension $d_c$ exists, such that
\begin{itemize}
\item for a system whose spatial dimension $d$ is smaller than $d_c$, discrete breathers of arbitrarily low 
energy can be found, i.e., no energy threshold exist,
\item for a system where $d\geq d_c$, an energy threshold exists, i.e., there is a positive lower bound on the 
energy of discrete breathers.
\end{itemize}

The existence or absence of an energy threshold for discrete breathers beautifully explains certain observations 
made in discrete systems. One example are statistical properties characterizing the spontaneous formation of 
discrete breathers in cooled lattices \cite{PiLeLi}, a second example are the power spectra observed in 
thermalized lattices \cite{ElFlaTsi}. In both cases, the presence of an energy threshold induces a qualitative 
change in the quantities under investigation.

The result on energy thresholds in \cite{FlaKlaMac} was derived under assumptions which turned out to be not 
general enough to cover all classes of systems of physical interest, and neither all of the different mechanisms 
leading to the (non)existence of an energy threshold. In this article, a much richer scenario is described, 
taking into account a larger class of systems and, most notably, pointing out three different mechanisms 
effecting the existence (or non-existence) of an energy threshold. The conditions leading to the occurrence of 
any of these mechanisms will be worked out, providing means to predict the existence of energy thresholds. A 
flavour of at least a part of these results has been given previously in a letter \cite{Kastner}. In the present 
article, a more general situation is considered and details and derivations are given.

There are three additional classes of systems, not covered in \cite{FlaKlaMac}, for which energy thresholds and 
critical dimensions are discussed in the present article: Firstly, systems with nonanalytic potential are 
included in the analysis. Albeit rare, physical systems with such interactions exist and are under experimental 
investigation, for example in the field of granular media \cite{Nesterenko}. Secondly, systems in which no low 
amplitude discrete breathers exist are considered. Even for the class of systems with analytic potentials, these 
systems constitute a generic subclass, which accounts for their possible relevance in physics. Thirdly, results 
indicating the absence of an energy threshold in systems with no linear spectrum are presented. These kind of 
systems have attracted much interest recently, and strongly nonlinear tunable phononic crystals showing such a 
behaviour are currently being constructed \cite{NestPrivComm}.

The article is organized as follows: in \sref{eqofmotion} some notation is fixed, defining the classes of systems 
under consideration. Then, it is appropriate to distinguish between systems {\em with}\/ and {\em without}\/ a 
linear spectrum. \Sref{phonons} is devoted to the first category, and the existence of an energy threshold is 
traced back to the occurrence of a certain modulational instability. Systems without a linear spectrum are 
treated in \sref{nophonons}. Finally, in \sref{summary}, the results for the various cases are summarized, 
providing a compilation of the different mechanisms which effect the existence of an energy threshold and 
specifying the conditions which give rise to each of these mechanisms.

\section{Equations of motion}\label{eqofmotion}

We consider a hypercubic lattice $\mathcal{L}_d=\{1,...,L\}^d$ in $d\in\mathbb{N}$ spatial dimensions. For 
simplicity, we consider $L\in 2\mathbb{Z}$ an even number. Each of the $N=L^d$ sites is labelled by a 
$d$-dimensional vector $n\in\mathcal{L}_d$, and to each site a state $(p_n,x_n)$ is assigned, where both, the 
momenta $p_n\in\mathbb{R}^f$ and the positions $x_n\in\mathbb{R}^f$, are vectors of a finite number $f$ of 
components. The dynamical properties of the system are gouverned by the Hamiltonian function
\begin{equation}\label{Hamil}
  H=\sum_{n\in\mathcal{L}_d} \left[ \frac{p_n^2}{2} + V(x_n) + \sum_{m\in \mathcal{N}_n} W(x_m-x_n) \right]
\end{equation}
with periodic boundary conditions in all spatial directions. $V$ is called on-site potential, $W\neq0$ 
interaction potential, and site $n$ is assumed to interact with its neighbourhood $\mathcal{N}_n$ of nearest 
neighbouring sites on the lattice.

The potentials $V$ and $W$ are both assumed to attain minima for zero argument, and, without loss of generality, 
$V(0)=0$ and $W(0)=0$. In contrast to previous related work \cite{Flach1,FlaKlaMac,DoFla}, analyticity (in the 
sense of the existence of a Taylor series) of the potentials around their minima is not required. By removing 
this restriction of analyticity, additional models of physical interest are included, as for example granular 
media are described by means of nonanalytic interaction potentials \cite{Nesterenko}.

Several of the above restrictions on the system are imposed only for notational simplicity of what follows. 
Extending the computations to systems of different lattice geometries or interactions not only with nearest 
neighbours should be possible without any substantial change and are discussed to some extend in \sref{general}. 
The case of long-range interactions is not included in the following analysis. A treatment of such systems can be 
found in \cite{Flach2}, and these results will be confronted with those for short-range interactions in 
\sref{general}.

For the sake of readability, analytic calculations will be performed for the one-dimensional case
\begin{equation}\label{H1d}
  H=\sum_{n=1}^N \left[ \frac{p_n^2}{2} + V(x_n) + W(x_{n+1}-x_n) \right],
\end{equation}
again with periodic boundary conditions $p_{n+N}=p_n$ and $x_{n+N}=x_n$, where momenta as well as positions are 
assumed to be one-component ($f=1$). The effect of the dimensionality on the results will be discussed in 
\sref{hidim}.

The existence of an energy threshold for discrete breathers will depend on the asymptotic behaviour of the 
potentials $V$ and $W$ close to their minima, and therefore we choose to represent $V$ and $W$ as power series in 
their arguments,
\begin{equation}\label{VW}
  V(x)=\sum_{\mu=0}^\infty \frac{v_\mu}{r_\mu}|x|^{r_\mu},\qquad W(x)=\sum_{\mu=0}^\infty 
\frac{\phi_\mu}{r_\mu}|x|^{r_\mu},
\end{equation}
with real (but not necessarily integer) exponents $r_\mu\geq1$. For convenience we consider the powers to be of 
increasing order, $r_\mu<r_{\mu+1}$ for all $\mu$. Some of the coefficients $v_\mu$ and/or $\phi_\mu$ may be 
zero, and in this way the expansions in \eref{VW} can also account for potentials $V$ and $W$ having different 
powers. Note that, by treating this class of potentials, we will be able to make statements about energy 
thresholds for the even larger class of potentials which have \eref{VW} as expansions around their minima. The 
potentials $V$ and $W$ in \eref{VW} are chosen to be symmetric, and this restriction is made to keep the 
calculations simpler and the presentation more readable, as the number of cases to be distinguished is reduced in 
this way. For systems with linear spectrum and in the case of analytic potentials, the analysis of systems with 
asymmetric potentials can be found in \cite{Flach1}. These results, as well as the possibility of a 
generalization of our results to more general systems with asymmetric potentials, are briefly discussed in 
\sref{general}.

The equations of motion following from \eref{H1d} are
\begin{equation}\label{eom}
  \ddot{x}_n + V'(x_n) + W'(x_n-x_{n-1}) - W'(x_{n+1}-x_n) =0,
\end{equation}
where a dot denotes a total derivative with respect to time, a prime the derivative of a function with respect to 
its argument. It is convenient to introduce a transform to normal coordinates
\begin{equation}\label{nc}
  Q_q=\frac{1}{N}\sum_{n=1}^N \rme^{\rmi q n}x_n,\qquad q=\frac{2\pi l}{N},\quad l\in\left\{{\textstyle 
-\frac{N}{2}+1,...\frac{N}{2}}\right\},
\end{equation}
where the inverse transform is given by
\begin{equation}
  x_n=\sum_q \rme^{-\rmi q n}Q_q.
\end{equation}
The normal coordinates $Q_q(t)$, like the original ones $x_n(t)$, are functions of time, but this dependence will 
often be suppressed in the following.
Rewriting the equations of motion \eref{eom} in terms of normal coordinates \eref{nc} yields
\begin{equation}\label{eomnc}
  \ddot{Q}_q + F_q(Q) = 0,
\end{equation}
where
\begin{eqnarray}
  \fl F_q(Q)=\frac{1}{N}\sum_{n=1}^N \rme^{\rmi q n} \left[ V'\left( \sum_{q'} \rme^{-\rmi q' n}Q_{q'} 
\right)\right.\nonumber\\
  \left. + W'\left( \sum_{q'} (1-\rme^{\rmi q'}) \rme^{-\rmi q' n}Q_{q'} \right) - W'\left( \sum_{q'} 
(\rme^{-\rmi q'}-1) \rme^{-\rmi q' n}Q_{q'} \right) \right]
\end{eqnarray}
and $Q$ denotes a vector with entries $Q_q$.

\section{Systems with linear spectrum}\label{phonons}

In this section, we want to refer to systems with interaction potential 
$W(x)=\frac{\phi_0}{2}x^2+\mathcal{O}(x^s)$ quadratic in leading order, i.e., $\phi_0\neq0$ and $s>2$. 
($\mathcal{O}$ denotes Landau's order symbol.) The on-site potential, in contrast, can be either zero ($V=0$), or 
of quadratic or higher order, $V(x)=\mathcal{O}(x^r)$ with $r\geq2$. In the notation of the expansions in 
\eref{VW} this means $r_0=2$ and $\phi_0\neq0$. This setting of parameters allows linearization of the equations 
of motion \eref{eom} with a non-trivial result,
\begin{equation}\label{leom}
  \ddot{Q}_q + \omega_q^2 Q_q = 0,
\end{equation}
where
\begin{equation}\label{linfreq}
  \omega_q^2 = v_0+4\phi_0 \sin^2 \left(\frac{q}{2}\right)
\end{equation}
denotes the frequency of the linear mode $q$. The set of all the $\omega_q^2$ is called the linear spectrum. The 
differential equations \eref{leom} obviously decouple, having simple sine waves as solutions. In the following we 
will be interested in solutions of the {\em nonlinear}\/ equations of motion \eref{eom} in the limit of small 
oscillation amplitudes in which the linearized equations of motion \eref{leom} are approached.

\subsection{Main ideas of this section}\label{ideas}

\begin{figure}[t]
  \begin{center}
    \includegraphics[width=6.4cm,height=10.78cm,angle=270]{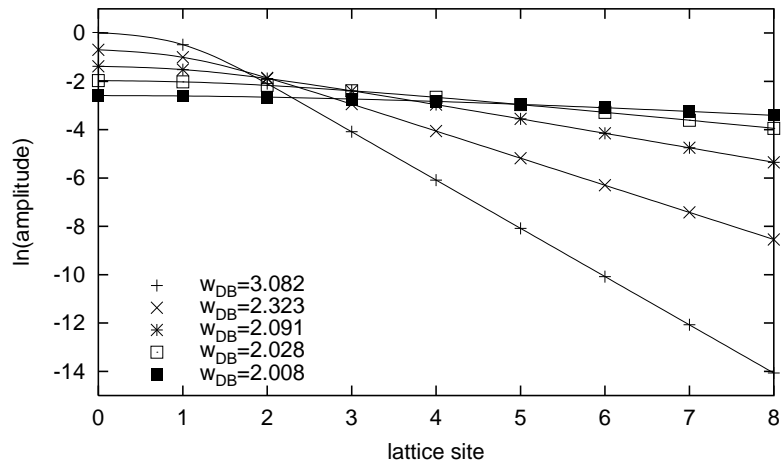}
  \end{center}
  \caption{\label{DBshape_el}Shape profiles (maximum values of the amplitude) of discrete breathers, centered at 
lattice site 0, in a Fermi-Pasta-Ulam chain of $N=99$ degrees of freedom (the equations of motion are given in 
\eref{FPUeom} with parameter value $a=0$). For a system which---like this one---has a linear spectrum, the 
localization is exponential (straight lines in the logarithmic plot). When the maximum breather amplitude is 
lowered (from + to $\blacksquare$), the localization strength decreases and the solution attains more and more 
the shape of a plane wave. Lines are merely drawn to guide the eye. 
}
\end{figure}

For generic Hamiltonian systems, discrete breathers, like all periodic orbits, occur in one-parameter families. 
Some typical choices of quantities to index such a family are the energy of a discrete breather, its frequency, 
or its amplitude measured at the site of maximum amplitude. In numerical studies of discrete breathers in a 
variety of systems with linear spectrum, the following scenario has been observed: consider some discrete 
breather with frequency $\omega_{\mbox{\tiny DB}}$ outside the linear spectrum. When following the family of 
discrete breathers and approaching the edge $\omega_{\mbox{\tiny edge}}$ of the band formed by the frequencies of 
the linear spectrum, it may or may not be the case that the breather amplitude $A$ goes to zero. The reverse 
conclusion, however, can be drawn: if discrete breathers of arbitrarily low amplitude exist, they cannot be found 
elsewhere but in the limit $\omega_{\mbox{\tiny DB}}\to\omega_{\mbox{\tiny edge}}$ \cite{FlaKlaMac}. Furthermore, 
in this case where
\begin{equation}\label{lowamp}
  \lim_{\omega_{\mbox{\tiny DB}}\to\omega_{\mbox{\tiny edge}}}A_{\mbox{\tiny DB}}(\omega_{\mbox{\tiny DB}})=0,
\end{equation}
it is observed numerically that in this limit the localization of the discrete breather becomes weaker and 
weaker. This fact is illustrated in \fref{DBshape_el}, where, for the example of the a Fermi-Pasta-Ulam chain  and 
discrete 
breathers of various frequencies, the maximum values of the amplitude are plotted for a number of lattice sites. 
More precisely, the localization strength approaches zero in the limit $\omega_{\mbox{\tiny 
DB}}\to\omega_{\mbox{\tiny edge}}$, and a transformation of a discrete breather into a spatially extended 
solution (plane wave) appears to take place. This observation gives rise to the following
{\hypothesis Consider a system with linear spectrum as specified above. Assume that in this system there exist 
low amplitude breathers, i.e., a family of discrete breathers for which \eref{lowamp} holds. Then low amplitude 
breathers emerge from band edge plane waves by means of a tangent bifurcation.}

A tangent bifurcation by definition takes place when a pair of Floquet multipliers of the linear stability 
analysis of the periodic orbit happens to collide at +1 on the unit circle. The periodic orbit emerging from the 
bifurcation appears to be of the same frequency as the one it originates from, and this is exactly the situation 
we observe for discrete breathers stemming from band edge modes.

The following three results or observations may serve to substantiate this hy\-poth\-e\-sis. Firstly, in a recent 
existence proof of discrete breathers due to James \cite{James2}, it is shown for a large class of 
one-dimensional systems that discrete breathers of small amplitude exist and that they emerge from a bifurcation 
of a band edge plane wave. Secondly, Flach has shown in \cite{Flach1} that the periodic orbits emerging from the 
tangent bifurcation lack a certain permutation symmetry. This is consistent with the periodic orbits being 
discrete breathers, but inconsistent with plane wave-type solutions. Thirdly, to the knowledge of the author, in 
the numerous numerical studies of discrete breathers, there has never been observed a family of discrete 
breathers in a system with linear spectrum which, in the limit (if it exists) of the amplitude going to zero, is 
not subject to the attenuation of the localization strength as described above.

The above hypothesis allows to tackle the problem of the existence of an energy threshold for discrete breathers 
in the following way (which is the outline of the rest of this section): instead of directly treating discrete 
breathers, the analytically more easily accessible band edge plane waves are considered. Since it is the 
behaviour at low amplitude which will be of importance, this can be done by employing perturbation techniques. 
Performing a Floquet analysis of these band edge plane waves, criteria for the appearance of a tangent 
bifurcation are found and the corresponding bifurcation energy is obtained. Then, the above hypothesis allows to 
relate this bifurcation energy to the energies of discrete breathers. In case the bifurcation energy is 
vanishing, no energy threshold is present. If the bifurcation energy is positive, an energy threshold has to be 
overcome in order to excite a discrete breather.

Note that there exist systems in which no low amplitude breathers are present and consequently the above 
hypothesis does not apply. This special case is deferred to \sref{nolowamp}, whereas for the following sections 
we assume \eref{lowamp} to be valid.

\subsection{Band edge modes}

At the band edges of the linear spectrum, there exist two particularly simple periodic orbits $Q^{\mbox{\tiny 
I}}$ and $Q^{\mbox{\tiny II}}$, each with only one mode excited,
\begin{equation}\label{bandedge}
  \begin{array}{r@{:\quad Q_q =0 \quad\forall q\neq}l@{\qquad\mbox{and}\qquad}r@{+V'(}c@{)}c@{\,=0,}}
    \rm{I}  & 0   & \ddot{Q}_0   & Q_0   &             \\\bs
    \rm{II} & \pi & \ddot{Q}_\pi & Q_\pi & +2W'(2Q_\pi)
  \end{array}
\end{equation}
where the excited mode is determined as the solution of the respective differential equation. $Q^{\mbox{\tiny 
I}}$ is called the in-phase mode, characterized by identical oscillations at all sites, i.e., $x_{n+1}(t)=x_n(t)$ 
for all $n\in\{1,...,N\}$ at all times $t$. As is obvious from the differential equation for $Q_0$ in 
\eref{bandedge}I, the presence of an on-site potential $V$ is essential for the existence of an in-phase mode. 
$Q^{\mbox{\tiny II}}$ we will refer to as the out-of-phase mode, where each two neighbouring sites oscillate in 
opposition of phase, i.e., $x_{n+1}(t)=-x_n(t)$ for all $n\in\{1,...,N\}$ at all times $t$. Exemplarily, the 
following computations will be presented for the in-phase mode only.

Of a uniformly valid perturbative solution for the differential equation \eref{bandedge}I, for our purposes the 
leading term in the energy $\varepsilon$ is sufficient,
\begin{equation}\label{oIsolution}
  Q_0(t)=\sqrt{\frac{2\varepsilon}{v_0}}\cos[\Omega(X)t+\varphi]+\mbox{h.o.t.},
\end{equation}
which can be obtained for example by means of a Lindstedt-Poincar\'e expansion \cite{Nayfeh} at low energy
\begin{equation}\label{defE}
  \varepsilon(Q^{\mbox{\tiny I}})=\frac{1}{2}\dot{Q}_{0}^2 + V(Q_0).
\end{equation}
$\varphi$ is an arbitrary phase shift which will be set to zero in the following, and higher order terms (h.o.t.) 
in the energy have been neglected. To render the expansion uniform, the nonlinear oscillation frequency
\begin{equation}\label{omega}  
\Omega(X)=\sqrt{v_0}\left[1+\frac{v_1}{v_0\sqrt{\pi}}\frac{\Gamma(\frac{r_1+1}{2})}{\Gamma(\frac{r_1+2}{2})}X 
+\mbox{h.o.t.}\right]
\end{equation}
has to be determined up to first order in
\begin{equation}
  X=\left(\frac{2\varepsilon}{v_0}\right)^{(r_1-2)/2},
\end{equation}
where $\Gamma$ is the Euler gamma function and higher order terms in $X$ have been neglected. This series 
expansion can be derived either by standard perturbation techniques \cite{Nayfeh} or using a method put forward 
in \cite{DoFla}.

\subsection{Tangent bifurcation}\label{tanbifu}

Our aim is to determine the energy at which a band edge mode undergoes a tangent bifurcation. Such a bifurcation 
implies a pair of Floquet multipliers of the linear stability analysis of the periodic orbit to collide at +1 on 
the unit circle, and is hence accompanied by a change in stability. In order to investigate the stability 
properties of the band edge modes, a small perturbation is considered by substituting
\begin{equation}
  Q_q \to Q_q + \delta_q
\end{equation}
in the equations of motion \eref{eomnc}. In the case of the periodic orbit $Q^{\mbox{\tiny I}}$, this 
substitution gives rise to equations of motions for the perturbations $\delta_q$ which, up to linear order in 
$\delta_q$, read
\begin{equation}\label{hill}
  \ddot{\delta}_q(t) + \left[ \mathcal{V}''(t) + \omega_q^2\right]\delta_q(t) = 0,
\end{equation}
where
\begin{equation}
  \mathcal{V}''(t)=V''[Q_0(t)]-v_0,
\end{equation}
and $\omega_q^2$ is defined in \eref{linfreq}. Note that this equation has a periodic coefficient 
$\mathcal{V}''$, as $Q_0$ is a periodic function in time. This type of differential equation (ordinary, 
homogeneous, second order, with periodic coefficient) is termed Hill equation, and a number of theorems 
concerning stability are at hand. Let us briefly recall some properties of the Hill equation (for details see 
\cite{MaWi}) and outline how they can be exploited in order to determine the bifurcation energy.

It follows from Floquet theory that, depending on the values of the parameters present in \eref{hill}, the Hill 
equation has either stable, unstable, or periodic solutions. Considering a suitable two-parameter space---in our 
case we choose the linear frequency $\omega_q$ and the energy $\varepsilon$, where the latter in turn determines 
$\mathcal{V}''$---the regions of stable solutions display a particular shape in parameter space, the so-called 
Arnold tongues (see figure 11-3 in \cite{Nayfeh} for an illustration). Regions of stable solutions are separated 
from regions of unstable solutions by lines of periodic solutions. Hence, the parameter values at which the band 
edge mode $Q^{\mbox{\tiny I}}$ bifurcates can be obtained by determining the values at which {\em any} of the 
$\delta_q$ has a periodic solution. The tangent bifurcation we are looking for appears at the smallest non-zero 
energy for which a periodic solution \eref{oIsolution} of orbit I and a periodic solution of the corresponding 
variational equation \eref{hill}, both with the same period, exist. In the following, by determining an energy 
expansion of the frequency $\tilde{\Omega}(\varepsilon)$ of the periodic solution of the variational equations 
\eref{hill}, and equaling it with the frequency $\Omega(\varepsilon)$ of the periodic orbit \eref{omega}, the 
bifurcation energy will be obtained to leading order.

The periodic solutions, as well as the corresponding parameter values, can be obtained by perturbation methods. 
In order to apply the Lindstedt-Poincar\'e technique \cite{Nayfeh}, we introduce the transformation 
$\tau=\tilde{\Omega}(X)t$ into \eref{hill}, yielding
\begin{equation}\label{hilltau}
 \tilde{\Omega}^2(X)\ddot{\delta}_q(X,\tau)+\left[\mathcal{V}''(X,\tau)+\omega_q^2\right]\delta_q(X,\tau)=0,
\end{equation}
where a dot now denotes a total derivative with respect to $\tau$. This change of variable transforms 
$\mathcal{V}''$ in a function which is $\pi$-periodic in $\tau$. The energy dependence has been noted explicitly 
by the variable $X$. Inserting the Ans\"atze
\begin{eqnarray}
  \delta_q(X,\tau) = \delta_{q,0}(\tau) + \delta_{q,1}(\tau)X + \mbox{h.o.t.},\\
  \tilde{\Omega}^2(X) = \tilde{\Omega}_0^2+\tilde{\Omega}_1^2 X + \mbox{h.o.t.},\label{Otilde}
\end{eqnarray}
as well as the series expansions \eref{oIsolution} and \eref{omega} in the Hill equation \eref{hilltau}, terms of 
equal order in $X$ can be collected to obtain the differential equations
\begin{eqnarray}
  \mbox{order}\;X^0:\quad & \tilde{\Omega}_0^2 \ddot{\delta}_{q,0}(\tau) + \omega_q^2 \delta_{q,0}(\tau) = 
0,\label{zeroO}\\
  \mbox{order}\;X^1:\quad & \tilde{\Omega}_0^2 \ddot{\delta}_{q,1}(\tau) + \omega_q^2 \delta_{q,1}(\tau) + 
\tilde{\Omega}_1^2 \ddot{\delta}_{q,0}(\tau) + \mathcal{V}_1''(\tau)\delta_{q,0}(\tau) = 0,\label{firstO}
\end{eqnarray}
where the $\pi$-periodic function
\begin{equation}
  \mathcal{V}_1''(\tau) = v_1(r_1-1)|\cos\tau|^{r_1-2}
\end{equation}
has been defined. The solution of the zeroth order equation \eref{zeroO} is
\begin{equation}\label{sol0}
  \delta_{q,0}(\tau) = A\cos\left(\frac{\omega_q \tau}{\tilde{\Omega}_0}\right) + B\sin\left(\frac{\omega_q 
\tau}{\tilde{\Omega}_0}\right),
\end{equation}
where $A$ and $B$ are complex valued constants. It follows from Floquet theory that, in order to obtain periodic 
solutions of period $2\pi$ of the first order equation \eref{firstO} in the low energy limit, it is sufficient to 
consider only the first two terms of a Fourier series of the driving term,
\begin{equation}
  \mathcal{V}_1''(\tau) = \sum_{k=0}^\infty C_k \cos(2k\tau)
\end{equation}
with coefficients
\begin{eqnarray}
  C_0 = \frac{2v_1}{\sqrt{\pi}}\frac{\Gamma(\frac{r_1+1}{2})}{\Gamma(\frac{r_1}{2})},\\
  C_1 = 2^{2-r_1}\frac{v_1(r_1-2)\Gamma(r_1)}{\Gamma(\frac{r_1+2}{2})\Gamma(\frac{r_1}{2})}.\label{coeffC1}
\end{eqnarray}
Inserting the truncated Fourier series as well as the zeroth order solution \eref{sol0}, the first order equation 
\eref{firstO} reads
\begin{equation}\label{diffFou}
  \tilde{\Omega}_0^2 \ddot{\delta}_{q,1}(\tau) + \omega_q^2\delta_{q,1}(\tau) + \left[ 
D+C_1\cos(2\tau)\right]\delta_{q,0}(\tau) = 0,
\end{equation}
where
\begin{equation}
  D=C_0-\left(\frac{\tilde{\Omega}_1\omega_q}{\tilde{\Omega}_0}\right)^2.
\end{equation}
The differential equation \eref{diffFou} is linear, and its solutions, which are in general aperiodic, can be 
obtained. Then, the periodic ones are found by eliminating the secular terms in the solution. This elimination is 
achieved only under certain conditions on the frequency, where the ones corresponding to the tangent bifurcation 
we are looking for are given by
\begin{eqnarray}
  \tilde{\Omega}_0^2=\omega_q^2,\\
  \tilde{\Omega}_1^2=C_0+\frac{C_1}{2}.
\end{eqnarray}
Equalling the truncated series expansions of the squared frequencies $\Omega^2$ and $\tilde{\Omega}^2$, and 
solving for the variable $X$, the critical values
\begin{equation}\label{Ebifuq}
  \varepsilon_{c,q}^{\mbox{\tiny 
I}}=\frac{v_0\vphantom{\omega_q^2}}{2}\left(\frac{v_0-\omega_q^2}{C_1}\right)^{2/(r_1-2)}
\end{equation}
of the energy at which a tangent bifurcation of orbit I may occur are obtained. Note that, in the preceding 
derivation of the bifurcation energy, we have simultaneously considered $N$ uncoupled equations \eref{hill} for 
the perturbations $\delta_q$, labelled by the index $q$. From $\delta_0$ no bifurcation can arise, as this 
component simply effects a time shift of the periodic orbit \cite{Flach1}. A change of stability in any of the 
remaining equations, however, may give rise to a bifurcation. We are interested in the tangent bifurcation giving 
the lowest value of the bifurcation energy \eref{Ebifuq}, which implies the choice $q=2\pi/N$. For a large number 
of lattice sites $N$, we can write the corresponding linear frequency \eref{linfreq} as
\begin{equation}
  \omega_{2\pi/N}^2 = v_0 + 4\pi^2\phi_0 N^{-2}+\mathcal{O}(N^{-4}).
\end{equation}
Inserting this expansion as well as equation \eref{coeffC1}, the bifurcation energy of orbit I reads
\begin{equation}\label{EcI}
  \varepsilon_c^{\mbox{\tiny I}} = \varepsilon_{c,2\pi/N}^{\mbox{\tiny I}} \simeq 2v_0\left[\frac{\pi^2\phi_0 r_1 
\Gamma(\frac{r_1-2}{2})\Gamma(\frac{r_1}{2})}{-N^2v_1\Gamma(r_1)}\right]^{2/(r_1-2)}
\end{equation}
in leading order of $N^{-1}$. Since $v_0,\phi_0>0$ and since we have to demand a positive bifurcation energy, the 
tangent bifurcation can only occur if
\begin{equation}\label{condI}
  v_1<0.
\end{equation}

A similar analysis for the out-of-phase mode (periodic orbit II) yields the expression
\begin{equation}\label{EcII}
  \varepsilon_c^{\mbox{\tiny II}} \simeq \frac{v_0+4\phi_0}{2}\left[\frac{\pi^{5/2} \phi_0 r_1 
\Gamma(\frac{r_1-2}{2})}{2 N^2(v_1+2^{r_1}\phi_1)\Gamma(\frac{r_1+1}{2})}\right]^{2/(r_1-2)}
\end{equation}
for the bifurcation energy. We have $v_0\geq0$ and $\phi_0>0$, leading to the necessary condition
\begin{equation}\label{condII}
  v_1+2^{r_1}\phi_1>0
\end{equation}
for a tangent bifurcation to take place. Similar results in \cite{Flach1}, derived for the case of analytic 
potentials $V$ and $W$, are recovered by setting $r_1=4$. It was already noted in the same reference that 
conditions \eref{condI} and \eref{condII} mirror the fact that, for a tangent bifurcation to take place, it is 
necessary that the frequency of the band edge mode is repelled from the linear spectrum of the system with 
increasing energy.

\subsection{\label{hidim}Higher spatial dimensions}

The computations and results of the preceding section can be generalized to systems of higher spatial dimensions 
$d$. The changes in the expressions of the bifurcation energies \eref{EcI} and \eref{EcII} are of small extent, 
and for our purposes it will only be important to note how the proportionality between the bifurcation energies 
$\varepsilon_c$ and the system size $N$ changes.

For arbitrary spatial dimension $d$, the normal coordinates \eref{nc} are labelled by a $d$ dimensional vector 
$q=(q_1,...,q_d)$ with $q_i=\frac{2\pi l_i}{L}$ and $l_i\in\{-L/2+1,...,L/2\}$ for all $i$. Then, choosing the 
value of $q$ giving rise to the lowest bifurcation energy, a factor $L^{4/(2-r_1)}$ is obtained in the 
expressions of the bifurcation energy instead of $N^{4/(2-r_1)}$. Since $N=L^d$, this is incorporated into 
\eref{EcI} and \eref{EcII} by replacing $N$ by $N^{1/d}$. In this way for both, the in-phase as well as the 
out-of-phase mode, the proportionality
\begin{equation}\label{Ecprop}
  \varepsilon_c\propto N^{4/[d(2-r_1)]},
\end{equation}
connecting the bifurcation energy $\varepsilon_c$ with the system size $N$, is obtained.

\subsection{Energy thresholds for discrete breathers}

The hypothesis of \sref{ideas} allows us to relate the bifurcation energy of the band edge plane wave to the 
minimum energy accessible for a discrete breather. It follows from \eref{nc} and \eref{defE} that the bifurcation 
energy determined above is an intensive quantity, i.e., on the scale of total energy per particle, and we will 
now turn to its extensive counterpart
\begin{equation}
  E_c = N\varepsilon_c.
\end{equation}
Considering the proportionality \eref{Ecprop} in the limit of infinite system size $N\to\infty$, it is found that
\begin{equation}\label{enthresh}
  \lim_{N\to\infty}E_c \propto \lim_{N\to\infty} N^{1-4/[d(r_1-2)]} \left\{
    \begin{array}{r@{0\quad\mbox{if}\quad d}c@{d_c}l}
      = & <    & ,\\
      > & \geq & ,
    \end{array}
  \right.
\end{equation}
with critical dimension
\begin{equation}\label{d_c}
  \textstyle d_c=\frac{4}{r_1-2}.
\end{equation}
Hence, for systems of spatial dimension $d$ smaller than the critical value $d_c$, discrete breathers of 
arbitrarily small energy can be found, whereas for systems where $d\geq d_c$ an energy threshold has to be 
overcome in order to excite a discrete breather. \Eref{d_c} is in perfect agreement with the critical dimension 
$d_c=2$ obtained in \cite{FlaKlaMac} for (generic) analytic potentials with a non-zero quadratic and quartic 
term, which, with our restriction to symmetric potentials, corresponds to a nonlinearity of leading order $r_1=4$.

This result for the critical dimension can be recovered by considering a continuum ap\-prox\-i\-ma\-tion of the 
discrete system as explained in \cite{FlaKlaMac}. In this approximation, the breather energy $E_b$ can expressed 
as an integral (equation~(6) of reference~\cite{FlaKlaMac}) which, in the limit of small breather amplitudes $A$, 
yields the proportionality $E_B\propto A^{(4-zd)/2}$. For our choice of symmetric potentials $V$ and $W$ in 
\eref{VW}, the detuning exponent $z$ (see \cite{FlaKlaMac} for a definition) is equal to $r_1-2$. Then, 
considering $E_b$ in the limit $A\to0$, the expression \eref{d_c} for the critical dimension $d_c$, 
distinguishing between finite and diverging limiting values for $E_b$, is reproduced.

\subsection{Intuitive understanding}

The mechanism which gives rise to an energy threshold for discrete breathers in high enough spatial dimension can 
be understood on an intuitive level. When lowering the amplitude of a discrete breather towards zero, two 
different mechanisms, competing with respect to their effect on the breather energy, take place: the localization 
strength of the discrete breather tends to become weaker, leading to an increase in energy, while the mere 
lowering in amplitude causes a decrease in energy (see \fref{DBshape_el} for an illustration). Depending on the 
respective strengths of these effects, an energy threshold may or may not exist. In the limit of weak 
localization, the degrees of freedom far from the breather's centre gain in importance. Their contribution to the 
energy depends on their 'number', which in turn depends on the spatial dimension of the system, and it is in this 
way that $d$ enters the game.

\subsection{Numerical confirmation of the results}\label{numerics}

Numerical computations have been performed in order to confirm the above result on energy thresholds for discrete 
breathers in systems with linear spectrum. By numerical continuation of periodic orbits from an anti-continuum 
limit \cite{MaAub}, discrete breathers on finite lattices can be computed numerically up to machine precision. In 
particular, it is possible to continue a discrete breather along its family while varying a parameter. In doing 
so for a set of frequencies approaching the band edge of the linear spectrum, the dependence of the breather 
energy on its amplitude, measured at the site of maximum amplitude, can be determined. In order to test our main 
result, the existence or non-existence of an energy threshold of discrete breathers under certain conditions, we 
will confront such data for two exemplary systems, one displaying an energy threshold, the other not.

\begin{figure}[t]
  \begin{center}
    \includegraphics[width=5.92cm,height=10.78cm,angle=270]{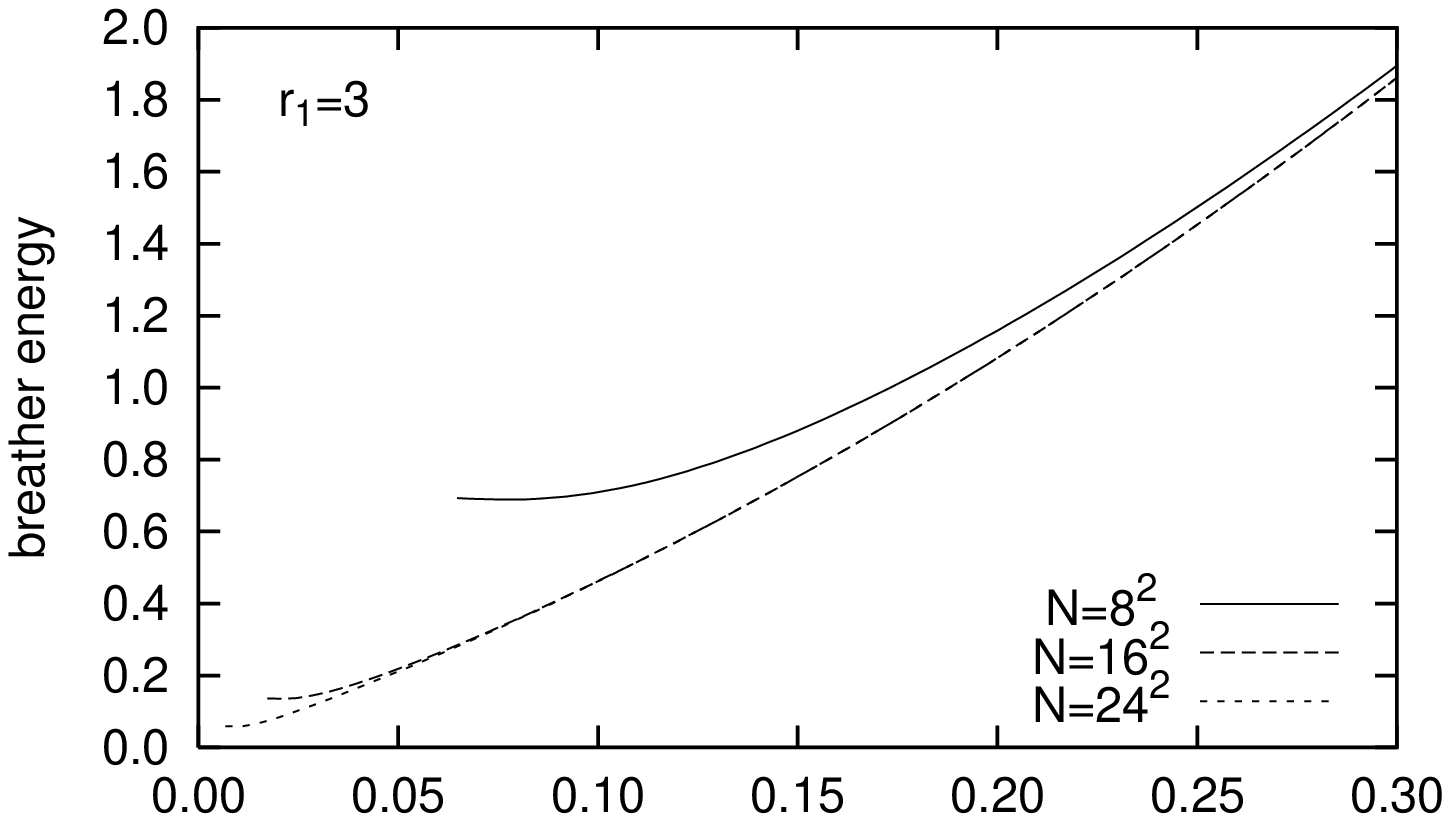}
    \includegraphics[width=6.4cm,height=10.78cm,angle=270]{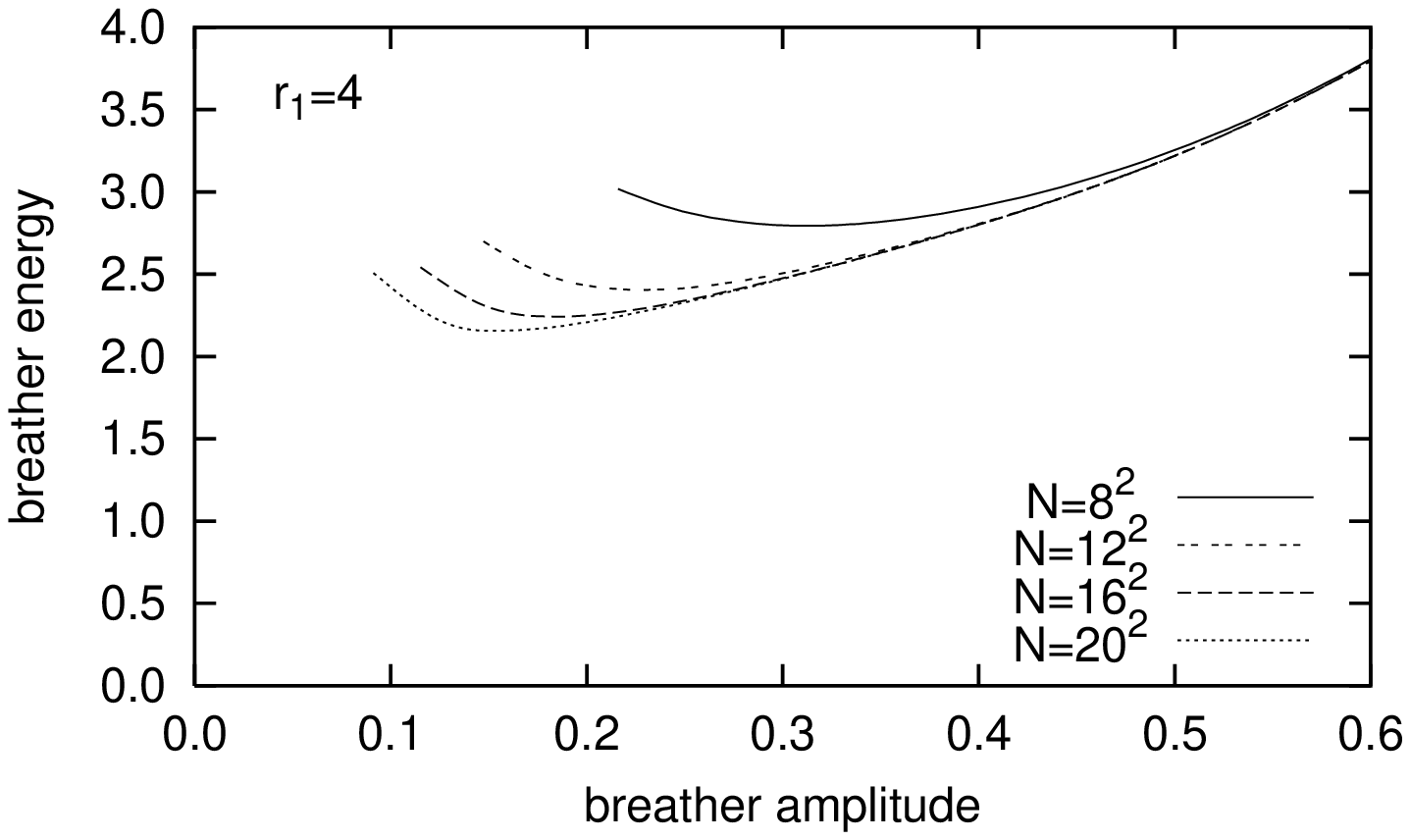}
  \end{center}
  \caption{\label{en_vs_amp}Breather energy versus amplitude for two-dimensional Fermi-Pasta-Ulam systems 
\eref{eq_motion} with periodic boundary conditions for the cases $r_1=3$ and $r_1=4$. A lower bound on the 
breather energy is observed. For $r_1=3$, with increasing system size $N$ this bound converges towards zero, 
whereas for $r_1=4$ it approaches a non-zero value.}
\end{figure}

Consider a two-dimensional system ($d=2$) of Fermi-Pasta-Ulam type, i.e., with zero on-site potential $V=0$. The 
interaction potential $W$ in the Hamiltonian \eref{Hamil} is chosen as
\begin{equation}
  W(x)=\frac{1}{2}x^2+\frac{1}{r_1}|x|^{r_1},
\end{equation}
giving rise to equations of motion of the form
\begin{equation}\label{eq_motion}
  \ddot{x}_n + \sum_{m\in \mathcal{N}_n}(x_n-x_m)\left[1+|x_n-x_m|^{r_1-2}\right] = 0,
\end{equation}
where $\mathcal{N}_n$ denotes the set of nearest neighbours of lattice site $n$. Then, following \eref{enthresh} 
and \eref{d_c}, no energy threshold is present for $r_1=3$, whereas for the choice of $r_1=4$ a threshold is 
obtained. These predictions are confirmed by the numerical results for finite systems with periodic boundary 
conditions presented in \fref{en_vs_amp}.
For any finite number of lattice sites, a lower bound on the breather energy is observed. For $r_1=3$, with 
increasing system size $N$ this bound converges towards zero, whereas for $r_1=4$ it approaches a non-zero value. 
Note that, for the system sizes considered, the range of validity of the large $N$ approximation \eref{EcII} has 
not yet been reached, as, in this approximation, no system size dependence of the energy threshold is expected in 
the case of $r_1=4$.

\subsection{Generalizing to larger classes of systems?}\label{general}

The above result on energy thresholds for discrete breathers has been obtained for the class of systems defined 
in \sref{eqofmotion}. However, several of the restrictions have been imposed merely to keep the presentation 
clearer and for notational convenience, and we will discuss in the following which ones presumably could be 
removed.

It is obvious from \eref{enthresh} that the existence of an energy threshold is ruled by the power of the system 
size $N$ in the expressions of the bifurcation energy \eref{EcI} and \eref{EcII}. This power, in turn, can be 
traced back to the powers of the energy $X$ in the series expansions of the frequency \eref{omega} of the band 
edge mode, and of the frequency \eref{Otilde} at which parametric resonance occurs. Although we have not 
explicitly done the calculations, we do not expect the powers in these expansions---and hence the energy 
thresholds for discrete breathers---to vary under a number of generalizations like
\begin{itemize}
\item other than hypercubic lattice geometries,
\item states $(p_n,x_n)$ with more than two components ($f>1$ in the notation of \sref{eqofmotion}),
\item interaction with a finite neighbourhood larger than nearest neighbours only,
\end{itemize}
Energy thresholds for discrete breathers in systems with Hamiltonian functions of other than the standard form 
\eref{Hamil}, like the discrete nonlinear Schr\"odinger equation, have been considered in \cite{FlaKlaMac} and, 
on a rigorous level (i.e., without invoking our hypothesis 1), in \cite{Weinstein}. Systems with asymmetric $V$ 
and $W$ are discussed in \cite{Flach1} for the case of analytic potentials. There it is found that allowing for 
asymmetric contributions in the series expansions \eref{VW} changes the expression for the bifurcation energy 
$E_c$, but it does not change the critical dimension $d_c$, and we would assume the same to happen for 
nonanalytic potentials. What {\em can} be modified by the inclusion of an asymmetric term, however, is whether 
low amplitude breathers do or do not exist in a certain system, and this issue is discussed in the next section.

For systems with long range interactions, the existence of energy thresholds is investigated in \cite{Flach2}. 
There it is found that the presence of long-range interactions enhances the appearance of an energy threshold, as 
shown for one-dimensional systems with analytic potentials, in which, in contrast to the short-range case, an 
energy threshold is observed.

\subsection{Systems without low amplitude breathers}\label{nolowamp}

The above analysis leading to our result on energy thresholds for discrete breathers is heavily relying on the 
hypothesis stated in \sref{ideas}, where it is assumed that breathers of arbitrarily low amplitude 
$A_{\mbox{\tiny DB}}$ exist. This condition is fulfilled in many cases, but exceptions not only do exist, but 
they even appear to be generic. The class of systems without low amplitude breathers is defined by the property
\begin{equation}\label{largeamp}
  \lim_{\omega_{\mbox{\tiny DB}}\to\omega_{\mbox{\tiny edge}}}A_{\mbox{\tiny DB}}(\omega_{\mbox{\tiny DB}})\neq0
\end{equation}
for all families of discrete breather present in the system when the breather frequency $\omega_{\mbox{\tiny 
DB}}$ approaches the frequency $\omega_{\mbox{\tiny edge}}$ of any of the band edge modes. First we want to give 
an example of a system without low amplitude breathers and study its characteristics. As a second step, some 
speculations are made on the general conditions leading to the absence of low amplitude breathers.

Consider a Fermi-Pasta-Ulam (FPU) chain, defined by Hamiltonian \eref{H1d} with
\begin{equation}
  V=0\qquad\mbox{and}\qquad W(x)=\frac{1}{2}x^2+\frac{a}{3}x^3+\frac{1}{4}x^4,
\end{equation}
which gives rise to the equations of motion
\begin{equation}\label{FPUeom}
  \ddot{x}_n + \sum_{m=n\pm1}[(x_n-x_m)+a(x_n-x_m)^2+(x_n-x_m)^3] = 0.
\end{equation}
For this system, two rigorous proofs of the existence of discrete breathers have been published. The first one is 
due to Aubry, Kopidakis and Kadelburg \cite{AuKoKa}, using a variational method. These authors proof existence of 
discrete breathers for any strictly convex interaction potential $W$, which corresponds to a parameter $a$ with 
$|a|<\sqrt{3}$ in \eref{FPUeom}. A second proof, applying a method of centre manifold reduction, was obtained by 
James \cite{James1}, where existence of discrete breathers has been shown for values of $a$ obeying 
$|a|<\frac{1}{2}\sqrt{3}$. Furthermore, as this latter method guarantees to catch all small amplitude solutions 
present in the system, the existence of small amplitude breathers can be excluded for $|a|>\frac{1}{2}\sqrt{3}$. 
Recapitulating, for the FPU chain \eref{FPUeom} and parameter values $\frac{1}{2}\sqrt{3}<a<\sqrt{3}$, existence 
of discrete breathers can be proofed, but small amplitude breathers in the sense of \eref{lowamp} are clearly 
absent. Knowing that low amplitude breathers do not exist, it is a triviality to infer the existence of an energy 
threshold for discrete breathers.

\begin{figure}[hbt]
  \begin{center}
    \includegraphics[width=6.4cm,height=10.78cm,angle=270]{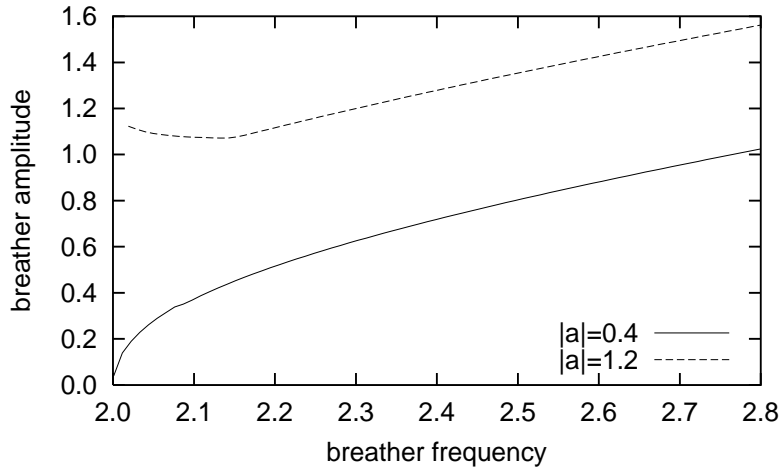}
  \end{center}
  \caption{\label{amp_vs_freq}Breather amplitude versus frequency for system \eref{FPUeom} with $N=99$ lattice 
sites, free boundary conditions, and frequencies close to the band edge of the linear spectrum (which is at 
frequency $2.0$). With parameter value $|a|=0.4<\frac{1}{2}\sqrt{3}$, breathers of arbitrarily low amplitude are 
observed, whereas for $\frac{1}{2}\sqrt{3}<|a|=1.2<\sqrt{3}$ the breather amplitude, and hence the breather 
energy, does not tend to zero when approaching the band edge of the linear spectrum.}
\end{figure}

To illustrate this result, numerical data, obtained by the continuation method described in \sref{numerics}, are 
presented. The two cases $a=0.4<\frac{1}{2}\sqrt{3}$ and $\frac{1}{2}\sqrt{3}<a=1.2<\sqrt{3}$ are considered, 
which allows to confront the behaviour of systems with and without low amplitude breathers. Varying the breather 
frequency $\omega_{\mbox{\tiny DB}}$ and approaching the band edge of the linear spectrum, the breather amplitude 
at the site of maximum amplitude is monitored. The result is presented in \fref{amp_vs_freq}, showing the 
amplitude to converge towards zero for $\omega_{\mbox{\tiny DB}}\to2$ in the first case ($a=0.4$), whereas a 
non-zero value is approached in this limit for the second case ($a=1.2$).

Some speculations can be made regarding the general conditions under which exclusively large amplitude breathers 
\eref{largeamp} are observed, implying the existence of an energy threshold for discrete breathers. At the end of 
\sref{tanbifu} we noticed already that, for a tangent bifurcation of the band edge mode to take place in the 
class of systems considered, the expansion coefficients of the potentials \eref{VW} have to fulfill the 
inequalities \eref{condI} and \eref{condII} for the in-phase and the out-of-phase mode, respectively. For the FPU 
system \eref{FPUeom}, the corresponding inequality was already obtained in \cite{Flach1}, equation (3.24), 
yielding $a\leq\frac{1}{2}\sqrt{3}$ in our notation. This inequality obviously coincides with the one deduced 
from the rigorous results \cite{AuKoKa} and \cite{James1}, identifying the values of $a$ providing low amplitude 
breathers. From this observation, one might deduce the following
{\hypothesis If, in a system with linear spectrum, no tangent bifurcation of any of the band edge modes take 
place, then discrete breathers of low amplitude in the sense of \eref{lowamp} do not exist.}

This hypothesis, if correct, allows to deduce the existence of an energy threshold for discrete breathers from 
the mere absence of a tangent bifurcation of the band edge modes. 

\section{Systems with no linear spectrum}\label{nophonons}

In analogy to the definition at the beginning of \sref{phonons}, by the expression ``systems with no linear 
spectrum'' we want to refer to systems whose potential functions $V$ and/or $W$ of the Hamiltonian \eref{Hamil} 
are not quadratic in leading order. In the notation of the expansions in \eref{VW} this means simply $r_0\neq2$. 
The peculiar properties of such systems have attracted much interest recently, and strongly nonlinear tunable 
phononic crystals showing such a behaviour are currently being constructed \cite{NestPrivComm}.

In contrast to systems with linear spectrum where discrete breathers are known to be exponentially localized in 
space \cite{MacAub,Flach3}, a localization even stronger than exponential is found in systems with no linear 
spectrum \cite{Flach3,DeyElFlaTsi,Yuan}. An upper bound on the breather amplitude is given by the inequality
\begin{equation}
  |x_i|\leq a\exp(-|i|/b)c^{(d^i/3)}
\end{equation}
for some $a,b,c,d\in\mathbb{R}$ in one-dimensional systems \cite{Yuan}, where the breather is centred at site 
$i=0$. However, arguments by which these results for the spatial decay can be obtained should be equally 
applicable in higher spatial dimension, and numerical computations confirm this reasoning, finding 
superexponential localization in higher dimensional systems.

\begin{figure}[htb]
  \begin{center}
    \includegraphics[width=6.4cm,height=10.78cm,angle=270]{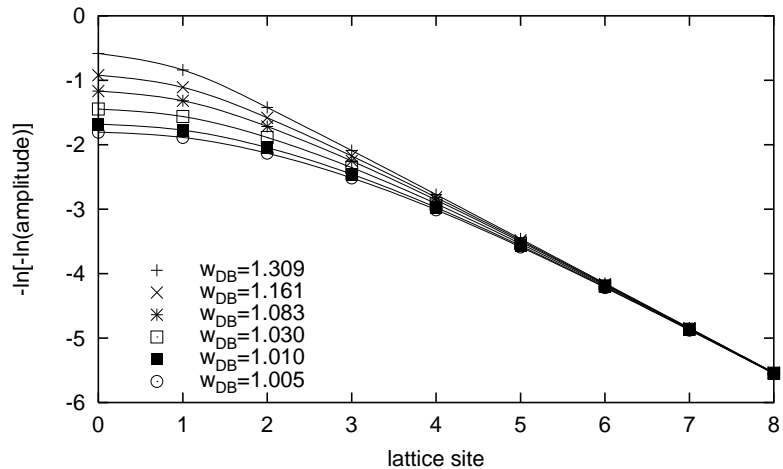}
  \end{center}
  \caption{\label{DBshape_sel}Shape profiles (maximum values of the amplitude) of discrete breathers, centered at 
site 0, in a system with no linear spectrum, namely the one-dimensional version of a Klein-Gordon type system with 
equations of motion \eref{KG} and $N=99$ degrees of freedom. The localization strength is superexponential and 
remains unchanged when varying the maximum breather amplitude (or, equivalently, the breather frequency 
$\omega_{\mbox{\tiny DB}}$. Note the twofold logarithmic scale of the ordinate 
($-\ln[-\ln(\cdot)]$). Lines are merely drawn to guide the eye.}
\end{figure}

Again, as in the preceding sections, we are interested in the low amplitude behaviour (if existing) of discrete 
breathers, which in turn affects the existence of an energy threshold. The mechanism as described in 
\sref{phonons}, the emergence of discrete breathers from a bifurcation of a band edge plane wave, cannot take 
place here simply due to the absence of a linear spectrum. Nevertheless, it might in principle be the case that, 
when lowering the amplitude of a discrete breather in a system with no linear spectrum, the localization strength 
decreases, giving rise to an energy threshold in some spatial dimension. For lack of any better idea, some 
numerics was performed, suggesting for the (few) cases considered that
\begin{enumerate}
\item \label{it1}low amplitude breathers appear to exist, and
\item \label{it2}the localization strength of discrete breathers remains constant in the low amplitude limit. 
\end{enumerate}
As an example, data are presented for a Klein-Gordon type system with Hamiltonian \eref{Hamil} and potentials
\begin{equation}
  V(x)=\frac{1}{2}x^2 \qquad\mbox{and}\qquad W(x)=\frac{1}{3}|x|^3,
\end{equation}
giving rise to the equations of motion
\begin{equation}\label{KG}
\ddot{x}_n = -x_n+\sum_{m\in \mathcal{N}_n}(x_m-x_n)|x_m-x_n|,
\end{equation}
where $\mathcal{N}_n$ denotes the set of nearest neighbours of lattice site $n$. Plotting the shape profiles of 
various discrete breathers obtained numerically for a one-dimensional system in \fref{DBshape_sel}, the above 
points \eref{it1} and \eref{it1} are corroborated.
Inspired by the numerical observations, we formulate the following
{\hypothesis Consider a system with no linear spectrum as specified above which supports discrete breathers. 
Then, low amplitude discrete breathers in the sense of \eref{lowamp} can be found. Furthermore, the localization 
strength is approximately constant within a family of discrete breathers.}

\begin{figure}[htb]
  \begin{center}
    \includegraphics[width=6.4cm,height=10.78cm,angle=270]{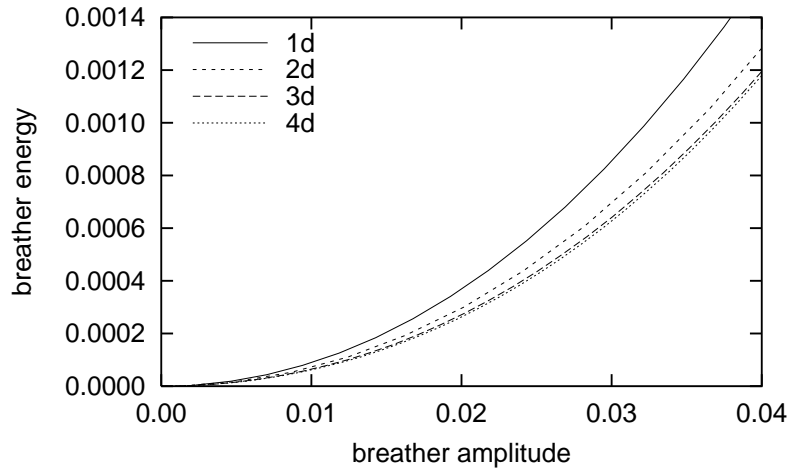}
  \end{center}
  \caption{\label{en_vs_amp03}Energy versus amplitude for discrete breath\-ers in a system with no linear 
spectrum. Independently of the spatial dimension $d=1,2,3,4$ (from top to bottom), arbitrarily low breather 
energies are observed. The data were obtained for systems consisting of $N=8^d$ lattice sites, but, due to the 
strong (superexponential) localization, are indistinguishable on this scale from those of larger systems.}
\end{figure}

Note that this hypothesis is based only on numerical observations in a few examples of systems, both of FPU and 
Klein-Gordon type. From this hypothesis, the absence of an energy threshold for discrete breathers in arbitrary 
spatial dimension is deduced immediately. For illustration, numerical data are presented in \fref{en_vs_amp03} 
for system \eref{KG} in spatial dimensions $d=1,2,3,4$. Absence of an energy threshold is observed in any of 
these dimensions already for small finite systems. This is clearly distinct from the behaviour observed in 
\fref{en_vs_amp} for systems with a linear spectrum where, if at all, the threshold vanishes in the limit of 
large system size only. 

\section{Summary of the results}\label{summary}

We have studied the existence of energy thresholds for discrete breathers in a large class of Hamiltonian 
systems. Breather energies are found to have a positive lower bound if the lattice dimension $d$ is greater than 
or equal to a certain critical value $d_c$, whereas no energy threshold is observed for $d<d_c$. The value of 
$d_c$ depends on the class of systems under consideration, and we can distinguish the following three cases:
\begin{enumerate}
\item Systems with continuous linear spectrum which support low amplitude breathers: discrete breathers of low 
amplitude emerge from a tangent bifurcation of a band edge mode. The bifurcation energy can be calculated 
explicitly and is related to a bound on the breather energy. For all finite systems, an energy threshold for 
discrete breathers is observed. For systems of large sizes $N$, this energy threshold scales as 
$N^{1-4/[d(r_1-2)]}$, resulting in a critical dimension $d_c=4/(r_1-2)$ for infinite systems. (See 
\sref{eqofmotion} for a definition of $r_1$.) 
\item Systems with linear spectrum which do not support low amplitude breathers: no tangent bifurcation of a band 
edge mode is observed, and apparently no other mechanism for the creation of low amplitude breathers exist. As an 
immediate consequence of the absence of low amplitude breathers, an energy threshold for discrete breathers 
exists for finite as well as infinite systems. 
\item Systems with no linear spectrum: the localization strength of discrete breathers does not vary 
significantly with the frequency, and strongly localized low amplitude breathers appear to exist. This brings 
about the absence of an energy threshold for discrete breathers for finite as well as infinite systems, i.e., 
$d_c=\infty$.
\end{enumerate}
Note that the above results are derived on the basis of hypotheses 1 to 3 formulated in sections \ref{phonons} 
and \ref{nophonons}. The hypotheses are corroborated in part by analytical results, in part by numerical 
observations.

\ack
Helpful comments from and discussions with J\'er\^ome Dorignac, Sergej Flach and Roberto Livi are gratefully 
acknowledged. Thanks to Oliver Schnetz for readily providing math\-e\-mat\-i\-cal support. This work was done 
during my stay at the Universit\`a di Firenze, Italy, in the group of Roberto Livi, supported by EU contract 
HPRN-CT-1999-00163 (LOCNET network).


\Bibliography{18}
\bibitem{MacAub} MacKay R S and Aubry S 1994 Proof of existence of breathers for time-reversible or Hamiltonian 
networks of weakly coupled oscillators \NL {\bf 7} 1623--43
\bibitem{Bambusi} Bambusi D 1996 Exponential stability of breathers in Hamiltonian networks of weakly coupled 
oscillators \NL {\bf 9} 433--57
\bibitem{LiSpiMac} Livi R, Spicci M and MacKay R S 1997 Breathers on a diatomic FPU chain \NL {\bf 10} 1421--34
\bibitem{AuKoKa} Aubry S, Kopidakis G and Kadelburg V 2001 Variational proof for hard discrete breathers in some 
classes of Hamiltonian dynamical systems {\it Discrete Contin.\ Dynam.\ Systems B}\/ {\bf 1} 271--98
\bibitem{James1} James G 2001 Existence of breathers on FPU lattices {\it C.\ R.\ Acad.\ Sci.\ Paris S\'er.\ I 
Math.}\/ {\bf 332} 581--6
\bibitem{James2} James G 2003 Centre manifold reduction for quasilinear discrete systems {\it J.\ Nonlinear Sci.} 
{\bf 13} 27--63
\bibitem{JaNo} James G and Noble P 2003 Breathers on diatomic FPU chains with arbitrary masses, in {\it 
Localization \& Energy Transfer in Nonlinear Systems} ed V\'azquez \etal (Singapore: World Scientific) pp~225--32 
\bibitem{Swanson_ea} Swanson B I, Brozik J A, Love S P, Strouse G F, Shreve A P, Bishop A R, Wang W-Z and Salkola 
M I 1999 Observation of intrinsically localized modes in a discrete low-dimensional material \PRL {\bf 82} 
3288--91
\bibitem{SchwarzEnSie} Schwarz U T, English L Q and Sievers A J 1999 Experimental generation and observation of 
intrinsic localized spin wave modes in an antiferromagnet \PRL {\bf 83} 223--6
\bibitem{TriMaOr}Tr\'{\i}as E, Mazo J J and Orlando T P 2000 Discrete breathers in nonlinear lattices: 
experimental detection in a Josephson array \PRL {\bf 84} 741-4
\bibitem{Binder_ea} Binder P, Abraimov D, Ustinov A V, Flach S and Zolotaryuk Y 2000 Observation of breathers in 
Josephson ladders \PRL {\bf 84} 745-8
\bibitem{EdHamm} Edler J and Hamm P 2002 Self-trapping of the amide I band in a peptide model crystal \JCP {\bf 
117} 2415--24
\bibitem{Mandelik_ea} Mandelik D, Eisenberg H S, Silberberg Y, Morandotti R and Aitchison J S 2003 Observation of 
mutually-trapped multi-band optical breathers in waveguide arrays \PRL {\bf 90} 253902 
\bibitem{Sato_ea} Sato M, Hubbard E, English L Q, Sievers A J, Ilic B, Czaplewski D A and Craighead H G 2003 
Study of intrinsic localized vibrational modes in micromechanical oscillator arrays {\it Chaos}\/ {\bf 13} 
702--15 
\bibitem{FlaKlaMac} Flach S, Kladko K and MacKay R S 1997 Energy thresholds for discrete breathers in one-, two-, 
and three-dimensional lattices \PRL {\bf 78} 1207--10
\bibitem{Flach2} Flach S 1998 Breathers on lattices with long range interaction \PR E {\bf 58} R4116--9
\bibitem{DoFla} Dorignac J and Flach S (unpublished)
\bibitem{PiLeLi} Piazza F, Lepri S and Livi R 2003 Cooling nonlinear lattices toward energy localization {\it 
Chaos}\/ {\bf 13} 637--45
\bibitem{ElFlaTsi} Eleftheriou M, Flach S and Tsironis G P 2003 Breathers in one-dimensional nonlinear 
thermalized lattice with an energy gap {\it Physica}\/ D {\bf 186} 20--6
\bibitem{Kastner} Kastner M 2003 Energy thresholds for discrete breathers \PRL {\bf 92} 104301
\bibitem{Nesterenko} Nesterenko V F 2001 {\it Dynamics of Heterogeneous Materials}\/ (Berlin: Springer)
\bibitem{NestPrivComm} Nesterenko V F private communication
\bibitem{Flach1} Flach S 1996 Tangent bifurcation of band edge plane waves, dynamical symmetry breaking and 
vibrational localization {\it Physica}\/ D {\bf 91} 223--43
\bibitem{Nayfeh} Nayfeh A H 1981 {\it Introduction to perturbation techniques}\/ (New York: Wiley)
\bibitem{MaWi} Magnus W and Winkler S 1966 {\it Hill's equation}\/ (New York: Wiley)
\bibitem{MaAub} Mar\'{\i}n J L and Aubry S 1996 Breathers in nonlinear lattices: numerical calculation from the 
anticontinuous limit \NL {\bf 9} 1501--28
\bibitem{Weinstein} Weinstein M I 1999 Excitation thresholds for nonlinear localized modes on lattices \NL {\bf 
12} 673--91
\bibitem{Flach3} Flach S 1994 Conditions on the existence of localized excitations in nonlinear discrete systems 
\PR E {\bf 50} 3134--42
\bibitem{DeyElFlaTsi} Dey B, Eleftheriou M, Flach S and Tsironis G P 2001 Shape profile of compactlike discrete 
breathers in nonlinear dispersive lattice systems \PR E {\bf 65} 017601 
\bibitem{Yuan} Yuan X 2002 Construction of quasi-periodic breathers via KAM technique {\it Comm.\ Math.\ Phys.} 
{\bf 226} 61--100
\endbib
\end{document}